\documentclass[12pt]{article}

\usepackage{braket}
\usepackage{mathrsfs}
 \usepackage{extpfeil}

\begin{document}

\begin{center}
{\LARGE{\bf{Generalized Weyl quantization on the cylinder and quantum phase. }}}
\end{center}

\bigskip\bigskip

\begin{center}
Maciej Przanowski \footnote{E-mail address:  maciej.przanowski@p.lodz.pl} and Przemys\l aw Brzykcy\footnote{E-mail address:  800289@edu.p.lodz.pl}
\end{center}

\begin{center}

{\sl  Institute of Physics, Technical University of  \L\'{o}d\'{z},\\ W\'{o}lcza\'{n}ska 219, 90-924 \L\'{o}d\'{z}, Poland.}\\
\medskip

\end{center}

\vskip 1.5cm
\centerline{\today}
\vskip 1.5cm

\begin{abstract}
Generalized Weyl quantization formalism for the cylindrical phase space $S^1 \times \mathbb{R}^1$ is developed. It is shown that the quantum observables relevant to the phase of linear harmonic oscillator or electromagnetic field can be represented within this formalism by the self-adjoint operators on the Hilbert space $L^2(S^1)$. 
\end{abstract}



 \section{Introduction} \label{sec1} 
The problem of defining the phase operator for a harmonic oscillator or for a single-mode electromagnetic field in quantum mechanics is an intriguing and still unsolved question. 
The existence of a self-adjoint phase operator $\hat{\Phi}$ canonically conjugate to the number operator~$\hat{N}$ 
\begin{equation}
\left[\hat{\Phi}, \hat{N} \right] =i
\label{1.1}
\end{equation}
was postulated by Dirac about 85 years ago \cite{1}.
However, in 1964 Susskind and Glogower \cite{2} showed that Dirac's assumption led to essential controversies (see also \cite{3,4}). In conclusion, instead of $\hat{\Phi}$ they have introduced the self-adjoint operators which can be interpreted as the cosine and sine operators of the phase. But this really interesting result has not closed the discussion as it seems quite clear that the well defined classical phase observable should have its quantum counterpart.
It is worth while to note that the problem with definition of $\hat{\Phi}$ as the self-adjoint operator canonically conjugate to $\hat{N}$ fulfilling (\ref{1.1}) can be easily understood as a direct consequence of the celebrated Pauli theorem \cite{5} under observation that $\hat{N}$ is bounded from below and its spectrum is discrete.
Recall that the same Pauli theorem causes severe difficulties with a correct definition of the time operator as the object canonically conjugate to the Hamilton operator \cite{6,7,8}. So, some methods applied to the problem of defining the time operator are analogous to the ones used in the case of searching for the quantum phase.
In particular, one can look for the phase operator by performing the Weyl quantization of the classical phase of harmonic oscillator considered as a function on the phase space $\mathbb{R}^2$ \cite{9}.
However, since this function is rather involved the corresponding operator obtained from the Weyl quantization rule can reveal properties which are not pertinent to the expected properties of the correct phase operator. The similar case occurs when the classical arrival time function is quantized \cite{7,10,11}. 
In 1970 Garrison and Wong \cite{3} were able to find the self-adjoint phase operator which satisfied the commutation relation (\ref{1.1}) on a dense subset of the Hilbert space (see also \cite{12,13}). The problem with the Garrison-Wong phase operator $\hat{\Phi}_{\mathrm{GW}}$ is that the probability distribution of the phase calculated 
for $\hat{\Phi}_{\mathrm{GW}}$ in any eigenstate of the number operator $\hat{N}$ is not uniform \cite{12} (see also section \ref{sec5} of the present paper).
Yet, another approach to the definition of quantum phase has been considered by Popov and Yarunin \cite{14,15}
and then developed by Pegg and Barnett \cite{16,17,18},and nowadays is called \textit{the Pegg-Barnett (PB) approach}. We will study it in more detail in our paper. Here we only point out that the main idea of the PB approach is to define the phase operator in the appropriate sequence of finite-dimensional Hilbert spaces. Then all calculations concerning a given observable relevant to the phase are first accomplished in those finite-dimensional Hilbert spaces and then one takes the limit with the dimension tending to infinity. 
Some objections against this approach has been raised by Busch, Grabowski and Lahti \cite{13}.
Namely, they write "Nevertheless there is no reason to stick to the finite-dimensional Hilbert space: one may equally well do all calculations after performing the limit $s\to\infty$"(\cite{13} p.6. Here $s$ stands for the dimension of respective Hilbert space).
In Ref. \cite{13} the quantum phase is given by a \textit{positive operator valued (POV) measure} (see also \cite{19}) and this POV measure leads to the Pegg-Barnett results but without any use of finite-dimensional Hilbert spaces. It is also proved in \cite{13} that POV measure defining the quantum phase arises from some spectral measure$E:\mathcal{B} ([-\pi,\pi) ) \to \mathcal{L}_+ (L^2(S^1)) $   by the Naimark projection $\hat{\Pi}:L^2(S^1) \to L^2(\mathbb{R}^1)$ (see section \ref{sec5} of the present paper). This result shows that the Hilbert space of states for a particle on the circle, $L^2(S^1)$, \textit{seems to play the crucial role for understanding the quantum phase}. The same conclusion follows from a nice work by Sharatchandra \cite{20}.
The aim of our paper is to develop this idea in more detail. 
We intend also to show how the PB approach can be incorporated into the generalized Weyl quantization formalism. 
In section \ref{sec2} we introduce the \textit{generalized Weyl application} and the \textit{generalized Stratonovich-Weyl (GSW) quantizer} for a particle on the circle.
In section \ref{sec3} we use the idea of the Pegg-Barnett approach to get the restricted GSW quantizer and to employ this quantizer in defining quantum observables on the cylindrical phase space $S^1\times\mathbb{R}^1$. 
The results of sections \ref{sec2} and \ref{sec3} enable us to find in section \ref{sec4} the angle operator with the use of GSW quantizer. We demonstrate that  one can apply the Pegg-Barnett approach to rotation angle observable in a "natural way" and this   leads to the sequence of angle operators in finite-dimensional Hilbert spaces quite different from the respective sequence obtained in \cite{21}. 
Section \ref{sec5} is devoted to the problem of incorporating the quantum phase into the generalized Weyl quantization strategy on $S^1\times\mathbb{R}^1$. 
Our proposition of the solution of this problem is described by the points $(1)$, $(2)$ and $(3)$ (see section \ref{sec5}).
As is then shown, this approach leads to the self-adjoint operator on the Hilbert space $L^2(S^1)$ which gives the same results as the POV measure approach of Refs. \cite{13,19} and the Pegg-Barnett approach \cite{16,17,18}.
Moreover, the analogous strategy can be used for other physical quantities which depend on the phase $\phi$ and/or the number $N$. For example, in section \ref{sec6} we use it to find the uncertainty relation for $\hat{\Phi}$ and $\hat{N}$.


\section{Generalized Stratonovich-Weyl quantizer for a particle on the circle. }\label{sec2}
Let the angle coordinate on the unit circle $S^1$ be denoted by $\Theta$, $-\pi \leqslant \Theta < \pi$. The Hilbert space $L^2(S^1)$ can be identified with $L([-\pi , \pi))$ or equivalently with $L^2(2\pi)$ which is the vector space of equivalence classes of the complex $2\pi$-periodic functions on $\mathbb{R}^1$ equipped with the scalar product $\braket{f|q} = \int_a^{a+2\pi} f^*(x)g(x) \, \mathrm{d}x$ , $\ket{f},\ket{g}  \in L^2(2\pi)$ and $a\in \mathbb{R}^1$ (see \cite{22} for details). The angle and the angular momentum operators are denoted by $\hat{\Theta }$ 
 and $\hat{L}$, respectively.
[$\hat{\Theta}$ is an operator in $L^2(S^1)$ which can be recognized as the multiplication by the angle $\Theta$ when the space $L^2(S^1)$ is identified with $L^2([-\pi , \pi ))$. Analogously the angular momentum operator $\hat{L}$ in $L^2(S^1)$ can be identified with the operator $-i\hslash\frac{\partial}{\partial \theta}$, $\theta \in (-\infty , \infty)$, 
acting in $L^2(2\pi)$]. As it is well known \cite{23} the angle operator $\hat{\Theta}$ is bounded and self-adjoint. Then $\hat{L}$ is unbounded self-adjoint operator of the domain $\mathscr{D}(\hat{L}) \in L^2(S^1)$ consisting of the classes of equivalence of absolutely continuous functions on $S^1$ such that 
\begin{equation}
\int _{-\pi} ^{\pi} \left| \frac{\mathrm{d} f(\Theta)}{\mathrm{d} \Theta} \right|^2 \,\mathrm{d} \Theta < \infty,
\end{equation}
Eigenvectors $\ket{l}, \, l\in\mathbb{Z}$ of $\hat{L}$ ($\braket{l|l'} = \delta_{ll'}$) can be identified
with the functions $ \frac{1}{\sqrt{2\pi}} \exp{(il\Theta)}$, $l\in\mathbb{Z}$. As usually one has 
\begin{equation}
\hat{L} \ket{l} =\hslash l \ket{l},\quad l\in\mathbb{Z}.\label{2.2}
\end{equation}
Given the self-adjoint operators $\hat{\Theta}$ and $\hat{L}$ we define   the unitary operators on $L^2(S^1)$ as $\exp{(in\hat{\Theta})}$, $n\in\mathbb{Z}$  and $\exp{\left( \frac{i}{\hslash}\alpha\hat{L}\right)}$, $\alpha\in\mathbb{R}^1$.
From the very definitions of $\hat{\Theta}$ and $\hat{L}$, employing also well known formulas of operator algebra one quickly finds the important commutation relation 
\begin{equation}
\left[ \hat{L} , \exp{\{in\hat{\Theta}\}} \right] := \hat{L} \exp{\{in\hat{\Theta}\}} -\exp{\{in\hat{\Theta}\}} \hat{L} = \hslash n \exp{\{in\hat{\Theta}\}}, \quad \forall{n\in\mathbb{Z}} \label{2.3} 
\end{equation}
and the following relation
\begin{equation}
\exp{\bigg\{ \frac{i}{\hslash}\alpha\hat{L}\bigg\}}\exp{\{in\hat{\Theta}\}} =  
\exp{\{i \alpha n\}} \exp{\{in\hat{\Theta}\}}  \exp{\bigg\{ \frac{i}{\hslash}\alpha\hat{L}\bigg\}}\, \quad \forall{n\in\mathbb{Z}, \alpha\in \mathbb{R}^1}.\label{2.4}
\end{equation}
Equations (\ref{2.2}) and (\ref{2.3}) yield 
\begin{equation}
\exp{\{i n \hat{\Theta}\}} \ket{l} = \ket{l+n}, \quad \forall{l,n\in \mathbb{Z}}. \label{2.5}
\end{equation}
The crucial role in quantization on the cylindrical phase space $S^1\ \times \mathbb{R}^1$ is played by the 
following unitary operator on $L^2(S^1)$  \cite{24,25,26,27}
\begin{eqnarray}
\hat{U}(\sigma,l) &=& \exp{\Big\{ i\left( \frac{\sigma}{\hslash} \hat{L} + l \hat{\Theta} \right) \Big\}}:=  \exp{\Big\{ \frac{i}{2} \sigma l \Big\} } \exp{\{il\hat{\Theta} \}} \exp{\Big\{ \frac{i}{\hslash} \sigma \hat{L} \Big\}} \stackrel{\mathrm{by}\, (\ref{2.4})}{=}\nonumber \\
&=& \exp{\Big\{ -\frac{i}{2} \sigma l \Big \} }  \exp{\Big\{ \frac{i}{\hslash} \sigma \hat{L} \Big\}} 
\exp{\{ il\hat{\Theta} \}}
\stackrel{\mathrm{by}\, (\ref{2.5}),(\ref{2.2})}{=}\nonumber \\
&=& \sum _{k=-\infty} ^ {\infty} \exp{\bigg\{ i \sigma \left( k+ \frac{l}{2} \right) \bigg \}} \ket{k+l} \bra{k} 
, \quad \sigma \in \mathbb{R}^1, l\in \mathbb{Z} \label{2.6} 
\end{eqnarray} 
With the use of the unitary operator $\hat{U}(\sigma, l)$ given by (\ref{2.6}) we can accomplish the quantization on $ S^1 \times \mathbb{R}^2$.  Examples of such a quantization have been studied in \cite{24,25,26,27}.
Here we contemplate the generalized version of Ref. \cite{27}. To this end we employ the results on generalized Weyl quantization for $\mathbb{R}^{2m}$ \cite{28,29,30,31,32}. So, if $f=f(\Theta,L)$ is a function on $S^1\times \mathbb{R}^1$ (the classical observable) then the corresponding operator $\hat{f}$ (the quantum observable) is given by the \textit{generalized Weyl application for the kernel $\mathcal{K}$} (see also \cite{11})  
\begin{equation}
\hat{f}\left[ \mathcal{K} \right] = \sum_{n=-\infty} ^{\infty} \int_{-\pi}^{\pi}
 f(\Theta,n\hslash) \hat{\Omega}\left[ \mathcal{K} \right](\Theta,n)
 \frac{\mathrm{d}\, \Theta}{2\pi}\label{2.7}
\end{equation}
where $\hat{\Omega}\left[ \mathcal{K} \right] (\Theta,n)$ is the \textit{generalized Stratonovich-Weyl (GSW) quantizer for the kernel $\mathcal{K}$}
\begin{equation}
\hat{\Omega}\left[ \mathcal{K} \right](\Theta,n):= \sum_{l=-\infty}^{\infty} \int_{-\pi}^{\pi} \mathcal{K}(\sigma,l) \exp{\{ -i(\sigma n + l\Theta)\}} \hat{U}(\sigma,l) 
\frac{\mathrm{d} \sigma}{2\pi}
\label{2.8}.
\end{equation}
The kernel $\mathcal{K} = \mathcal{K}(\sigma,\lambda)$, $\sigma\in[-\pi, \pi ] $, $\lambda\in\mathbb{R}^1$, is a smooth function with respect to $(\sigma,\lambda) $. 
Analogously as in the case of phase space $\mathbb{R}^{2m}$ also in the present case the kernel $\mathcal{K}$ determines the ordering of operators and one can also show that the "natural" assumptions about the properties of the correspondence (\ref{2.7}) impose some restrictions on the function $\mathcal{K}(\sigma,\lambda)$. 
Thus, for example, we get 
\begin{itemize}
\item
{
   The operator $\hat{f}[\mathcal{K}]$ is symmetric for any real classical observable $f=f(\Theta,L)$ iff 
   \begin{equation}
     \forall \sigma \in[-\pi,\pi],     l\in\mathbb{Z} \quad   \mathcal{K}^*(-\sigma,-l) = \mathcal{K}(\sigma,l)\label{2.9},
   \end{equation}
}
\item
{
$\hat{f}[\mathcal{K}] = f(\hat{\Theta})$ for any function $f$ depending only on $\Theta$, $f=f(\Theta)$, iff 
\begin{equation}
 \forall l\in \mathbb{Z} \quad \mathcal{K}(0,l) = 1, 
\label{2.10}
\end{equation}
}
\item
{
$\hat{f}[\mathcal{K}] = f(\hat{L})$ for any function f depending only on $L$, $f=f(L)$, iff 
\begin{equation}
\forall \sigma\in[-\pi,\pi]\quad\mathcal{K}(\sigma,0) = 1.
\label{2.11}
\end{equation}
}
\end{itemize} 
It is an easy matter to demonstrate that the property (\ref{2.9}) ensures that $\hat{\Omega}[\mathcal{K}]$ is a symmetric operator, and the property (\ref{2.11}) yields 
\begin{equation}
\mathrm{Tr}  \big\{ \hat{\Omega}[\mathcal{K}](\Theta,n) \big\}=1.
\label{2.12}
\end{equation}
Moreover, if one imposes the following "natural" condition on the form of the operator $\hat{A}_{j,k}$, $j,k\geq 0$, corresponding to the monomial $A_{j,k} = L^j\Theta^k $ according to the prescription (\ref{2.7}):
\begin{equation}
\hat{A}_{j,k} = \sum_{s=0} ^ {\min{(j,k)}} g(j,k,s) \hslash^s \hat{L}^{j-s} \hat{\Theta}^{k-s}, \quad g(j,k,s)\in\mathbb{C}
\label{2.13}
\end{equation}
then, assuming also that $\mathcal{K}(\sigma,\lambda)$ is an analytic function, one concludes that $\mathcal{K}$ is a function of the variable $\sigma\lambda$ (see the analogous considerations in \cite{31}). In this case we will simply write 
\begin{equation}
\mathcal{K} = \mathcal{K}(\sigma\lambda).
\label{2.14}
\end{equation}
Note that with (\ref{2.14}) assumed the conditions (\ref{2.10}) and (\ref{2.11}) are satisfied iff
\begin{equation}
\mathcal{K}(0)=1
\label{2.15}
\end{equation}
and the condition (\ref{2.9}) holds iff $\mathcal{K}(\sigma\lambda)$ is a real function
\begin{equation}
\mathcal{K}^* = \mathcal{K}
\label{2.16}.
\end{equation}
\subsection{Examples}
\begin{itemize}
\item[\textit{(i)}]{\textit{Weyl ordering.}}

Here we assume
\begin{equation}
\mathcal{K}=1\label{2.17}.
\end{equation}
This case has been analyzed at length in \cite{27}. In particular it has been shown that GSW quantizer reads now 
\begin{eqnarray}
&&\hat{\Omega}[1](\Theta,n)  =  \exp{\{ -2in\Theta\}} \sum_{k=-\infty}^{\infty} \exp{\{2ik\Theta\}} 
 \bigg[
\ket{2n-k} \bra{k}  
\nonumber \\  &+& \frac{2}{\pi} 
\sum_{l=-\infty}^{\infty}  \frac{(-1)^l}{2l+1} \exp{ \{ -i(2l+1) \Theta \} }
\ket{2(n+l) - k +1 } \bra{k} 
 \bigg]  \label{2.18}
\end{eqnarray}
\item[\textit{(ii)}]{\textit{Symmetric ordering.}}

This case was utilized in our previous paper \cite{11} on the arrival time operator for a particle on a circle.
We put now 
\begin{equation}
\mathcal{K} = \cos{\left( \frac{\sigma\lambda}{2} \right)}
\label{2.19}
\end{equation}
(compare with \cite{31} ). Inserting (\ref{2.19}) into (\ref{2.8}) and carrying out  straightforward calculations one gets the respective GSW quantizer in the following form 
\begin{eqnarray}
\hat{\Omega}\left[\cos{\left( \frac{\sigma\lambda}{2} \right)}\right](\Theta,n) &=& \frac{1}{2} \sum_{k=-\infty}^{\infty} \bigg[  \exp{\{-i(n-k)\Theta\}} 
\ket{n} \bra{k}  
\nonumber \\  &+&   \exp{\{-i(n-k)\Theta\}}  \ket{k} \bra{n}  
 \bigg]  \label{2.20}
\end{eqnarray}
\end{itemize}
A quick glance at (\ref{2.18}) or (\ref{2.20}) reveals that those formulas  and, consequently, also the main definition of GSW quantizer (\ref{2.8}), are fairly formal since they do not represent any operator. 
The analogous problem we find in the case when the GSW quantizer on $\mathbb{R}^{2m}$ is defined. 
As it is known in this last case the GSW quantizer  is an operator valued distribution rather than "usual" operator. 
In the present case we propose the procedure which follows from the ideas developed in \cite{14,15,16,17,18}, \cite{21} and \cite{33} in connection with investigations on the phase operator and rotation angle operator. We trust that our approach gives a new insight into those questions.


\section{Restricted GSW quantizer}\label{sec3}
Consider a $(2N+1)$-dimensional Hilbert space 
\begin{equation}
\mathcal{H}_N := \mathrm{span}\{ \ket{k} \}_{k=-N}^N \subset  L^2(S^1).
\label{3.1}
\end{equation}
Using the definition (\ref{2.6}) we can   restrict the operator $\hat{U}(\sigma,l)$ to the following operator acting on $\mathcal{H}_N$
\begin{equation}
\hat{U}_N(\sigma,l) := \sum_{\substack{-N \leqslant k \leqslant N \\ -N \leqslant k+l \leqslant N}}
\exp{ \big\{  
i\sigma \left( k+ \frac{l}{2} \right)
\big\}}
\ket{k+l} \bra{k} \label{3.2}
\end{equation}
where $\sigma \in \mathbb{R}^1$, $-N \leqslant l \leqslant N$. 
It is evident that $\hat{U}_N(\sigma, l)$ maps $\mathcal{H}_N$ into $\mathcal{H}_N$, $\hat{U}_N(\sigma\l): \mathcal{H}_N \to \mathcal{H}_N$; however, $\hat{U}_N(\sigma,l)$ is unitary on $\mathcal{H}_N$ iff $l=0$.
With the use of $\hat{U}_N(\sigma, l)$, and (\ref{2.8}) one defines the \textit{restricted GSW quantizer} 
$\hat{\Omega}_N [ \mathcal{K} ] (\Theta, n) : \mathcal{H}_N \to \mathcal{H}_N$, $-N\leqslant n \leqslant N$ by
\begin{equation}
 \hat{\Omega}_N [ \mathcal{K} ] (\Theta, n) := \sum_{l=-N}^{N} \int_{-\pi}^{\pi}  \mathcal{K} (\sigma,l) \exp{ \{ -i(\sigma n +l \Theta) \}} \hat{U}_N(\sigma,l) \frac{\mathrm{d}\, \sigma}{2\pi}
\label{3.3}
\end{equation}
where $-N \leqslant n \leqslant N$.
Substituting (\ref{3.2}) into (\ref{3.3}) and performing simple manipulations we get the formula 
\begin{eqnarray}
\hat{\Omega}_N[\mathcal{K}] (\Theta,n) &=& \sum_{j=-N}^N \sum_{k=-N}^N  
\exp{ \{ -i(j-k) \Theta\}  }\cdot \nonumber \\ 
&&\left(
\int_{-\pi}^{\pi} \mathcal{K}(\sigma,j-k) 
\exp{\bigg\{  i\sigma \bigg(\frac{j+k}{2}-n \bigg)   \bigg\}}
\frac{\mathrm{d}\, \sigma}{2\pi}
\right)\ket{j}\bra{k}, \label{3.4}
\end{eqnarray}
where $-N\leqslant n \leqslant N$. 
The following two properties of $\hat{\Omega}_N[\mathcal{K}](\Theta,n) : \mathcal{H}_N \to \mathcal{H}_N $ can be easily shown 
\begin{itemize}
\item{}
if (\ref{2.9}) is assumed then $\hat{\Omega}_N[\mathcal{K}](\Theta,n)$ is a symmetric operator
\begin{equation}
\hat{\Omega}^+_N[\mathcal{K}](\Theta,n)=\hat{\Omega}_N[\mathcal{K}](\Theta,n) \label{3.5}
\end{equation}
\item{}
if (\ref{2.11}) is satisfied then 
\begin{equation}
\mathrm{Tr} \{ \hat{\Omega}_N[\mathcal{K}](\Theta,n) \} =1.\label{3.6}
\end{equation}
\end{itemize}
Given $\hat{\Omega}_N$ and employing (\ref{2.7}) we define \textit{restricted generalized Weyl application} $f\mapsto \hat{f}_N[\mathcal{K}] : \mathcal{H}_N \to \mathcal{H}_N$ as
\begin{eqnarray}
\hat{f}_N[\mathcal{K}] &:=& \sum_{n=-N}^N \int_{-\pi}^{\pi} f(\Theta,n\hslash) \hat{\Omega}_N[\mathcal{K}](\Theta,n) \frac{\mathrm{d}\, \Theta}{2\pi}=\nonumber \\
&\stackrel{\mathrm{by}\, (\ref{3.4})}{=}& \sum^N_{\substack{n,k,l=-N \\ -N \leqslant k+l \leqslant N}} 
\left(  \int_{-\pi}^{\pi}   f(\Theta,n\hslash) \exp{\{  -il\Theta \} }  \frac{\mathrm{d}\, \Theta}{2\pi}  \right) \cdot\nonumber \\
&\cdot&
\left( 
\int_{-\pi}^{\pi} \mathcal{K}(\sigma,l) \exp{ \bigg\{ i\sigma \bigg(k+\frac{l}{2} -n\bigg) \bigg\} } \frac{\mathrm{d}\, \sigma}{2\pi}
\right) \ket{k+l}\bra{k}=\nonumber \\
&=& \sum^N_ {n,j,k=-N } 
\left(  \int_{-\pi}^{\pi}   f(\Theta,n\hslash) \exp{\{  -i(j-k)\Theta \} }  \frac{\mathrm{d}\, \Theta}{2\pi}  \right) \cdot\nonumber \\
&\cdot&
\left( 
\int_{-\pi}^{\pi} \mathcal{K}(\sigma,j-k) \exp{ \bigg\{ i\sigma \bigg(\frac{j+k}{2} -n\bigg) \bigg\} } \frac{\mathrm{d}\, \sigma}{2\pi}
\right) \ket{j}\bra{k} .
\label{3.7}
\end{eqnarray}
Note the following important properties of $\hat{f}_N[\mathcal{K}]$: 
\begin{itemize}
\item{
If (\ref{2.9}) holds true then for any real function $f=f(\Theta,L)$ the respective operator $\hat{f}_N[\mathcal{K}] : \mathcal{H}_N \to \mathcal{H}_N$ is symmetric i.e.
\begin{equation}
\hat{f}^+_N[\mathcal{K}] =\hat{f}_N[\mathcal{K}] \label{3.8}
\end{equation}
on $\mathcal{H}_N$.}
\item{
If (\ref{2.11}) is fulfilled then  
\begin{equation}
\mathrm{Tr}\{ \hat{f}_N[\mathcal{K}] \} = \frac{1}{2\pi} \sum_{n=-N}^N \int_{-\pi}^{\pi} f(\Theta,n\hslash) \mathrm{d} \, \Theta,
\label{3.9}
\end{equation}
and the operator $\hat{f}_n[\mathcal{K}]$ corresponding  to the unity function $f=1$ is he unity operator on $\mathcal{H}_N$ i.e.
\begin{equation}
\sum_{n=-N}^N \int_{-\pi}^{\pi} \hat{\Omega}_N[\mathcal{K}](\Theta,n) \frac{\mathrm{d}\, \Theta}{2\pi}=\sum_{n=-N}^N \ket{n} \bra{n} =: \hat{1}_{\mathcal{H}_N}. \label{3.10}
\end{equation}
}
\end{itemize}
\subsection{Examples}
\begin{itemize}
\item[\textit{(i')}]{\textit{Weyl ordering.}

Substituting (\ref{2.17}) into (\ref{3.4}) after some simple manipulations we obtain
\begin{eqnarray}
\hat{\Omega}_N[1](\Theta,n) &=& \exp{\{-2in\Theta \}} \bigg[  \sum^N_{\substack{-N \leqslant k \leqslant N \\ 2n-N \leqslant k \leqslant 2n+ N}}  exp{\{2ik\Theta \} } \ket{2n-k} \bra{k} \nonumber \\
&+& \frac{2}{\pi}   \sum^N_{\substack{-N \leqslant k \leqslant N \\ -N-2n \leqslant 2l+1-k \leqslant N-2n}}
\frac{(-1)^l}{2l+1} \exp{\{2ik\Theta\} } \exp{\{ -i(2l+1)\Theta} \cdot \nonumber \\
&\cdot& \ket{2(l+n)+1-k} \bra{k} \bigg], \quad -N\leqslant n\leqslant N 
\label{3.11}
\end{eqnarray}
}
\item[\textit{(ii')}]{\textit{Symmetric ordering.}

Inserting (\ref{2.19}) into (\ref{3.4}) one gets a quite pretty formula 
\begin{eqnarray}
&&\hat{\Omega}_N\left[ \cos{\frac{\sigma\lambda}{2}} \right] (\Theta,n) = \frac{1}{2} \sum_{k=-N}^N \bigg[ \exp{\{ -i(n-k) \Theta\}} \ket{n}\bra{k}+ \nonumber \\
&&+  \exp{\{ i(n-k) \Theta\}} \ket{k}\bra{n}\bigg], \quad -N\leq n \leq N
\label{3.12}
\end{eqnarray}}
\end{itemize}
Observe that the restricted GSW quantizers (\ref{3.11}) and (\ref{3.12}) can be formally obtained by projecting (\ref{2.18}) and (\ref{2.20}), respectively, on the Hilbert space $\mathcal{H}_N$. Thus, in general one can formally write 
\begin{equation}
\hat{\Omega}_N[\mathcal{K}](\Theta,n) = \hat{\mathcal{P}}_N \hat{\Omega}_N[\mathcal{K}] (\Theta,n) \hat{\mathcal{P}}_N, \quad -N\leqslant n \leqslant N
\label{3.13}
\end{equation}
where $\hat{\mathcal{P}}_N := \sum_{l=-N}^N \ket{l}\bra{l} : L^2(S^1) \to \mathrm{span}\{\ket{k}\}_{k=-N}^N$.
Coming back to the restricted generalized Weyl application (\ref{3.7}) we can in a natural way extend the operator $\hat{f}_N[\mathcal{K}] : \mathcal{H}_N \to \mathcal{H}_N$ to the operator $\hat{f}_N^{\mathrm(ext)}[\mathcal{K}] : L^2(S^1) \to L^2(S^1)$ by putting 
\begin{equation}
\hat{f}_N^{\mathrm(ext)}[\mathcal{K}] (\ket{\psi})= \left\{
\begin{array}{rl}
\hat{f}_N [\mathcal{K}] (\ket{\psi}) & \text{if } \ket{\psi}\in\mathcal{H}_N,\\
0 & \text{if } \ket{\psi}\in L^2(S^1) \ominus  \mathcal{H}_N.
\end{array} \right. \label{3.14}
\end{equation}
From (\ref{3.7}) and (\ref{3.14}) one easily finds the matrix representation of $\hat{f}_N^{\mathrm(ext)}[\mathcal{K}] $
\begin{eqnarray}
\braket{j|\hat{f}_N^{(\mathrm{ext})}|k}&=& \sum_{n=-N}^N \bigg[ \bigg( \int_{-\pi}^{\pi} f(\Theta, n\hslash)
\exp{\{-i(j-k)\Theta \}} \frac{\mathrm{d}\, \Theta}{2\pi}\bigg)\cdot \nonumber \\
&\cdot& \bigg( \int_{-\pi}^{\pi} \mathcal{K}(\sigma,j-k) 
\exp{\bigg\{ i\sigma \bigg( \frac{j+k}{2} - n \bigg)\bigg\}}
\frac{\mathrm{d}\, \sigma}{ 2\pi}
 \bigg) \bigg],\nonumber \\
&&\text{ for } |j|\leqslant N \text{ and } |k| \leqslant N, \nonumber \\
\braket{j|\hat{f}_N^{(\mathrm{ext})}|k}&=&0,\nonumber \\
&& \text{ for }|j|>N \text{ or } |k| > N. 
\label{3.15}
\end{eqnarray}
Consequently, the problem of finding the operator $\hat{f}[\mathcal{K}]$ defined by (\ref{2.7}) can be stated as follows: Find the operator $\hat{f}[\mathcal{K}]$ in $L^2(S^1)$ such that its matrix representation is given by
\begin{eqnarray}
\braket{j|\hat{f}[\mathcal{K}]|k}&=&\lim_{N\to\infty}\braket{j|\hat{f}_N^{\mathrm{(ext)}}[\mathcal{K}]|k} \stackrel{\mathrm{by}\, (\ref{3.15})}{=}\nonumber \\
&=&
\lim_{N\to\infty} \sum_{n=-N}^N \bigg[ \bigg( \int_{-\pi}^{\pi} f(\Theta,n\hslash) \exp{\{-i(j-k) \Theta \}} \frac{\mathrm{d}\, \Theta}{2\pi}\bigg) \cdot\nonumber \\
&\cdot& \int_{-\pi}^{\pi} \mathcal{K}(\sigma,j-k) 
\exp{ \bigg\{ i\sigma \bigg( \frac{j+k}{2} - n \bigg) \bigg \}}\frac{\mathrm{d}\, \sigma}{2\pi} \bigg], \quad j,k\in \mathbb{Z}.
\label{3.16}
\end{eqnarray}
Moreover, we want this operator to be self-adjoint for any real function $f=f(\Theta,L)$. This problem belongs to classical problems of functional analysis and in particular it concerns the questions of extending a given symmetric operator to the self-adjoint operator \cite{23}. We do not deal with the general case, but in the next section we consider in detail the angle operator. Substituting (\ref{2.17}) or (\ref{2.19})  into (\ref{3.15}) one easily obtains the matrix elements for the cases of Weyl and symmetric orderings: 
\begin{itemize}
\item[\textit{(i'')}]{\textit{Weyl ordering.}

\begin{eqnarray}
\braket{j|\hat{f}_N^{\mathrm{(ext)}}[1]|k} &=& \int_{-\pi}^{\pi} f\left(\Theta,\frac{j+k}{2}\hslash\right)
\exp{\{-i(j-k)\Theta\} } \frac{\mathrm{d}\, \Theta}{2\pi},\nonumber \\
&&\text{when} |j|, |k| \leqslant N \text{ and} j+k \text{ is an even number}, \nonumber \\
\braket{j|\hat{f}_N^{\mathrm{(ext)}}[1]|k} &=&\frac{2}{\pi} \sum_{n=-N}^N \bigg[ 
\frac{(-1)^{\frac{j+k-1}{2}-n}}{j+k-2n}
 \int_{-\pi}^{\pi} f\left(\Theta,n\hslash\right)
\exp{\{-i(j-k)\Theta\} } \frac{\mathrm{d}\, \Theta}{2\pi}\bigg],\nonumber \\
&& \text{when} |j|, |k| \leqslant N \text{ and }  j+k  \text{ is an odd number}, \nonumber \\
 \braket{j|\hat{f}_N^{\mathrm{(ext)}}[1]|k} &=&0,\nonumber \\
&& \text{ when } |j| > N  \text{ or } |k| > N  
 \label{3.17}
\end{eqnarray}
}
\item[\textit{(ii'')}]{\textit{Symmetric ordering.}

\begin{eqnarray}
\braket{j| \hat{f}_N^{\mathrm{(ext)}} \bigg[ \cos{\frac{\sigma\lambda}{2}} \bigg]|k}&=&
\int_{-\pi}^{\pi} \frac{f(\Theta,j\hslash)+f(\Theta,k\hslash)}{2}
\exp{\{-i(j-k)\Theta\}} 
\frac{\mathrm{d}\, \Theta}{2\pi},\nonumber \\
&& \text{ for } |j|, |k| \leqslant N, \nonumber \\
\braket{j| \hat{f}_N^{\mathrm{(ext)}} \bigg[ \cos{\frac{\sigma\lambda}{2}} \bigg]|k}&=&0,\nonumber \\
 &&\text{for } |j|>N \text{ or } |k|>N.
\label{3.18}
\end{eqnarray}
 }
\end{itemize}
Note that in the case of symmetric ordering the respective formula (\ref{3.18}) is quite simple.
The form of $\hat{f}_N  \bigg[ \cos{\frac{\sigma\lambda}{2}} \bigg]$ is also simple 
\begin{equation}
\hat{f}_N  \bigg[ \cos{\frac{\sigma\lambda}{2}} \bigg]=
\sum_{j,k=-N}^N\left(\int_{-\pi}^{\pi} \frac{f(\Theta,j\hslash)+f(\Theta,k\hslash)}{2}
\exp{\{-i(j-k)\Theta\}}
\frac{\mathrm{d}\, \Theta}{2\pi} \right)\ket{j}\bra{k} \label{3.19}.
\end{equation}
Now we are at the position where a generalization of the Pegg-Barnett ideas \cite{16,17,18,21} can be applied.
Namely, instead of the operator $\hat{f}[\mathcal{K}]$ corresponding to the classical observable $f=f(\Theta,L)$ one considers the sequence of operators $\{\hat{f}_N[\mathcal{K}] \}_{N=0}^{\infty}$ as the object representing the respective quantum observable. Each operator $\hat{f}_N[\mathcal{K}]$ acts in the finite-dimensional Hilbert space 
$\mathcal{H}_N:=\mathrm{span}\{\ket{k}\}_{k=-N}^N$, $\hat{f}_N: \mathcal{H}_N \to \mathcal{H}_N$. Of course $\mathrm{dim} \mathcal{H}_N=2N+1$. Then following the Pegg-Barnett approach to the problems of phase operator \cite{16,17,18} and of the angle operator \cite{21} we propose to calculate measurable quantities relevant to the quantum observable represented by the sequence  $\{\hat{f}_N[\mathcal{K}] \}_{N=0}^{\infty}$ in the finite-dimensional Hilbert spaces $\mathcal{H}_N$ first and after these calculations were done we let $N$ tend to $\infty$. In the next section we explore this idea in more detail for rotation angle. 


\section{The angle operator}\label{sec4}
We are going to study the case when 
\begin{equation}
f(\Theta,L)=\Theta.\label{4.1}
\end{equation}
Substituting (\ref{4.1}) into (\ref{3.7}) one quickly gets
 \begin{eqnarray}
\hat{\Theta}_N[\mathcal{K} ] &=& \sum_{n,j,k=-N}^{N} \bigg( \int_{-\pi}^{\pi} \Theta \exp{\{-i(j-k)\Theta \} }
\frac{\mathrm{d}\, \Theta}{2\pi} \bigg)\cdot \nonumber \\
&\cdot& \bigg( \int_{-\pi}^{\pi} \mathcal{K}(\sigma,j-k) 
\exp{\bigg\{ i\sigma \bigg(\frac{j+k}{2} -n \bigg) \bigg\}} \frac{\mathrm{d}\, \sigma}{2\pi} \bigg) 
\ket{j}\bra{k} = \nonumber \\
&=&
\sum^N_{\substack{n,j,k=-N \\ j \neq k}} 
\bigg( \frac{i(-1)^{j-k}}{j-k} 
\int_{-\pi}^{\pi} \mathcal{K}(\sigma,j-k) 
\exp{\bigg\{ i \sigma \bigg( \frac{j+k}{2} -n \bigg)\bigg\}  } \frac{\mathrm{d}\, \sigma}{2\pi} \bigg) \ket{j}\bra{k}
.\nonumber \\ \label{4.2}
\end{eqnarray}
Performing summation over $n$ we can write (\ref{4.2}) in the following form 
\begin{eqnarray}
\hat{\Theta}_N[\mathcal{K} ] &=& 
i\sum^N_{\substack{j,k=-N \\ j \neq k}} 
\bigg( \frac{(-1)^{j-k}}{j-k} 
\int_{-\pi}^{\pi} \mathcal{K}(\sigma,j-k) 
\frac{\sin{  [ \sigma (N+\frac{1}{2} ) ]}}{\sin{\frac{\sigma}{2}}}
\exp{\bigg\{ i \sigma   \frac{j+k}{2}  \bigg\}  } 
\frac{\mathrm{d}\, \sigma}{2\pi} 
\bigg) \ket{j}\bra{k}
.\nonumber \\ \label{4.3}
\end{eqnarray}
The matrix elements of $\hat{\Theta}_N[\mathcal{K}] \ : \mathcal{H}_N \to \mathcal{H}_N$ read
\begin{eqnarray} 
\braket{j| \hat{\Theta}_N[\mathcal{K}]|k} &=& i \frac{(-1)^{j-k}}{j-k} \int_{-\pi}^{\pi} \mathcal{K}(\sigma,j-k) 
\frac{\sin{  [ \sigma (N+\frac{1}{2} ) ]}}{\sin{\frac{\sigma}{2}}}
\exp{\bigg\{ i \sigma   \frac{j+k}{2}  \bigg\}  } 
\frac{\mathrm{d}\, \sigma}{2\pi}, \nonumber \\
&&|j|,|k|\leqslant N, \quad j\neq k,\nonumber \\
\braket{j| \hat{\Theta}_N[\mathcal{K}]|j}&=&0, \quad |j| \leqslant N
\label {4.4}.
\end{eqnarray} 
Since
\begin{eqnarray}
\lim_{N\to \infty} \frac{1}{2\pi} \frac{\sin{  [ \sigma (N+\frac{1}{2} ) ]}}{\sin{\frac{\sigma}{2}}} &=& \frac{1}{2\pi}
\sum_{n=-\infty}^{\infty} \exp{\{i\sigma n \}}= \sum_{k=-\infty}^{\infty} \delta(\sigma+2k\pi) =: \delta^{(S^1)}(\sigma)\nonumber \\
\label{4.5}
\end{eqnarray}
assuming also that (\ref{2.10}) holds true we conclude that according to the prescription (\ref{3.16}) the angle operator $\hat{\Theta}[\mathcal{K}]$ should be determined by the following matrix representation
 \begin{equation}
\braket{j|\hat{\Theta} [\mathcal{K}]|k}=\lim_{N\to\infty}  \braket{j|\hat{\Theta}_N^{\mathrm{(ext)}} [\mathcal{K}]|k}=\left\{
\begin{array}{rl}
i \frac{(-1)^{j-k}}{j-k} & \text{if } j \neq k,\\
0 & \text{if }j=k,
\end{array} \right. \label{4.6}
\end{equation}
$j,k\in \mathbb{Z}$. Therefore the angle operator is independent of the kernel $\mathcal{K}$ and it reads 
\begin{equation}
\hat{\Theta} = i \sum^{\infty} _{\substack{j,k=-\infty \\ j \neq k}}  \frac{(-1)^{j-k}}{j-k} \ket{j}\bra{k}
\label{4.7}.
\end{equation}
$\hat{\Theta}$ is a bounded self-adjoint operator defined on the all HIlbert space $L^2(S^1)$ of the norm 
\begin{equation}
\| \hat{\Theta}\| _{L^2(S^1)}=\frac{\pi}{\sqrt{3}}.
\label{4.8}
\end{equation}
Under the identification 
\begin{eqnarray}
\ket{n} &\longleftrightarrow& \frac{1}{\sqrt{2\pi}}\exp{\{ in \Theta \}}, \nonumber \\
L^2(S^1)\ni \ket{\psi}=\sum_{k=-\infty}^{\infty} c_k\ket{k} &\longleftrightarrow& \sum_{k=-\infty}^{\infty} c_k   \frac{1}{\sqrt{2\pi}}\exp{\{ ik \Theta \}} \in L^2([-\pi,\pi))
\label{4.9}
\end{eqnarray}
the angle operator  in the Schr\"{o}dinger representation  $\hat{\Theta}_S$ reads
\begin{equation}
\hat{\Theta}_S \psi(\Theta) = \Theta \psi(\Theta), \quad \Theta \in [-\pi,\pi).
\label{4.10}
\end{equation}
Concluding, if the condition (\ref{2.10}) is fulfilled then one arrives at the angle operator $\hat{\Theta}$  which is independent of the kernel $\mathcal{K}$. This $\hat{\Theta}$ is given by (\ref{4.7}). However, as can be seen from (\ref{4.2}),(\ref{4.3}) and (\ref{4.4})
the sequence $\{ \hat{\Theta}_N[\mathcal{K}]  \}_{N=0}^{\infty}$
depends on $\mathcal{K}$.

\subsection{Examples}
\begin{itemize}
\item[\textit{(i''')}]{\textit{Weyl ordering.}

From (\ref{3.17}) with $f(\Theta,n\hslash) = \Theta$  one easily finds 
\begin{eqnarray}
\braket{j|\hat{\Theta}_N[1]|k}&=& \frac{i}{j-k},\nonumber \\
&&\text{for } |j|, |k| \leqslant N, j\neq k, j+k \text { is an  even number,} \nonumber \\
\braket{j|\hat{\Theta}_N[1]|j}&=&0\nonumber \\
&&\text{for } |j|\leqslant N, \nonumber \\
\braket{j|\hat{\Theta}_N[1]|k}&=& -\frac{i}{j-k} \frac{2}{\pi} \sum_{n=-N}^N \frac{(-1)^{\frac{j+k-1}{2}-n}}{j+k-2n}  \nonumber \\
&&\text{for } |j|, |k| \leqslant N,  j+k \text { is an odd number.}  
\label{4.11}
\end{eqnarray}
}
\item[(\textit{ii''')}]{\textit{Symmetric ordering.}

Here we have
\begin{eqnarray}
\braket{j|\hat{\Theta}_N\bigg[\cos{\frac{\sigma\lambda}{2}}\bigg]|k}&=& i \frac{(-1)^{j-k}}{j-k},\quad |j|, |k| \leqslant N, \, j \neq k, \nonumber \\
\braket{j|\hat{\Theta}_N\bigg[\cos{\frac{\sigma\lambda}{2}}\bigg]|k}&=& 0, \quad|j| \leqslant N,
\label{4.12}
\end{eqnarray}}
or in the operator form 
\begin{equation}
\hat{\Theta}_N\bigg[\cos{\frac{\sigma\lambda}{2}}\bigg] = i \sum^N_{\substack{j,k=-N \\ j \neq k}} 
\frac{(-1)^{j-k}}{j-k} \ket{j} \bra{k}.
\label{4.13}
\end{equation}
\end{itemize}
We end this section with the observation that our formula (\ref{4.3}) corresponds to Eq. (3.14) of \cite{21}. 
Note also that from the point of view of the generalized Weyl quantization rule on the cylindrical phase space $S^1\times \mathbb{R}^1$ the most natural form of the sequence $\{ \hat{\Theta} _N [\mathcal{K} ]\}_{N=0} ^ {\infty}$ seems to be that given by (\ref{4.13}) for the symmetric ordering. Of course the respective sequence given in \cite{21} is quite different.


\section{From the angle operator to the phase operator}\label{sec5}
The problem of introducing to quantum theory the observable corresponding to the phase of harmonic oscillator (or to the phase of electromagnetic field) was first considered by Dirac in his famous work on quantization of electromagnetic field \cite{1}.  In his work Dirac assumed the existence of   a self-adjoint phase operator   canonically conjugate to the number operator.  However, Susskind and Glogower \cite{2} have proved that such an assumption leads to contradictions. Consequently, instead of the phase operator they have introduced the self-adjoint operators which can be interpreted as the cosine and sine operators of the phase.  Six years later, in 1970, Garrison and Wong \cite{3} succeeded in defining a self-adjoint phase operator. This operator can be written in the following form 
\begin{equation}
\hat{\Phi}_{GW} = \Phi_0 +\pi + \sum_{\substack{j,k \geq 0\\ j\neq k}}\frac{\exp{\{i(j-k)\Phi_0}\}}{i(j-k)} \ket{\underline{j}} \bra{\underline{k}}
\label{5.1}
\end{equation}
where $\Phi_0$ is arbitrary and $ \ket{\underline{j}}$, $ \ket{\underline{k}}$, $j,k=0,1,2,\dots$, denote the eigenvectors of number operator $\hat{N}$ 
\begin{eqnarray}
\hat{N}  \ket{\underline{j}}&=& j \ket{\underline{j}} \nonumber , \\
\braket{\underline{j}|\underline{k} } &=& \delta_{jk} \label{5.2}.
\end{eqnarray}
We underline the eigenvectors $\ket{\underline{n}}$  of the number operator $\hat{N}$ to distinguish them from the eigenvectors of $\hat{L}$ denoted by $\ket{j}$.
 The Garrison-Wong (GW) phase operator (\ref{5.1}) was also found by Popov and Yarumin \cite{14,15} and many others  (see e.g. \cite{12,13}).
Another, inspiring and elegant approach to the problem of defining quantum phase was developed by Pegg and Barnett \cite{15,16,17}.  The main idea of the Pegg-Barnett (PB) method lies in considering the sequence of finite dimensional Hilbert spaces  $\{\{\mathrm{span}\{ \ket{\underline{n}} \}_{n=0}^{s}  \}_{s=0}^{\infty}$. Of course, $\mathrm{dim} \,\mathrm{span}\{ \ket{\underline{n}} \}_{n=0}^{s} =s+1$. Then in each of the spaces one selects the \textit{reference phase state} 
\begin{equation}
\ket{\underline{\Phi_0}} = \frac{1}{\sqrt{s+1}} \sum_{n=0}^{s} \exp{\{in\Phi_0 \}} \ket{\underline{n}}
\label{5.3}
\end{equation}
and the remaining \textit{phase states} have the form
\begin{eqnarray}
\ket{\underline{\Phi_m}} &=& \frac{1}{\sqrt{s+1}} \sum_{n=0}^{s} \exp{\{in\Phi_m \}} \ket{\underline{n}},\quad m=0,1,\dots,s\label{5.4}\\
\braket{\underline{\Phi_m}|\underline{\Phi_l }} &=& \delta_{ml} ,\quad\quad\quad\quad\quad\quad\quad\quad\quad\quad\quad m,l=0,1,\dots,s. \nonumber 
\end{eqnarray}
From (\ref{5.4}) we get
\begin{equation}
\Phi_m = \phi_0 + \frac{2m\pi} {s+1}, \quad m=0,1,\dots s. \label{5.5}
\end{equation}
Finally the \textit{PB phase operator} in the Hilbert space $\mathrm{span}\{ \ket{\underline{n}} \}_{n=0}^{s}$ is given by its spectral decomposition 
\begin{eqnarray}
\hat{\Phi}_{PB}^{(s)} &:=& \sum_{m=0}^{s} \Phi_m \ket{\underline{\Phi_m} } \bra{\underline{\Phi_m}} =\nonumber \\
&=& \Phi_0 + \frac{s\pi}{s+1} +\frac{2\pi}{s+1}  \sum_{\substack{j,k=0 \\ j\neq k}}^{s} \frac{\exp{\{ i(j-k)\Phi_0\}}}{\exp{\{ i \frac{2\pi(j-k)}{s+1}\}}-1} \ket{\underline{j}} \bra{\underline{k}} \label{5.6}.
\end{eqnarray}
The analogous results have been obtained by Popov and Yarunin \cite{14,15} who have also proved that the sequence $\{ \hat{\Phi}_{PB}^{(s)} \}_{s=0}^{\infty}$ is weakly convergent to the GW phase operator (\ref{5.1}) for $s\to\infty$.
From this result one might draw the seemingly final conclusion that the quantum phase is simply represented by the GW phase operator. However, since in general $ \braket{\underline{\Psi} |f(\hat{\Phi}_{GW})|\underline{\Psi}} \neq \lim_{s\to\infty} \braket{\underline{\Psi} |f(\hat{\Phi}_{PB}^{(s)})|\underline{\Psi}}$ another approach is also possible and this is exactly what has been proposed by Pegg and Barnett. Namely, for any quantum observable relevant to the phase we first perform all calculations in the finite-dimensional Hilbert space $\mathrm{span}\{ \ket{\underline{n}} \}_{n=0}^{s}$, $s=0,1,\dots$ and after that we let $s$ tend to infinity. As it has been demonstrated in \cite{12} the GW phase operator and the PB approach lead to essentially different results. In particular the variance of phase $(\Delta \Phi_{GW})^2$ in the number eigenstate $\ket{\underline{n}}$ calculated directly using the GW phase operator (\ref{5.1}) reads
\begin{eqnarray}
(\Delta \Phi_{GW})^2 &=& \frac{\pi^2}{6}, \quad\quad\quad\quad \,\,\, \mathrm{for}\,n=0\nonumber \\
(\Delta \Phi_{GW})^2 &=&\frac{\pi^2}{6}+\sum_{k=1}^{n} \frac{1}{k},\quad \mathrm{for}\, n\geq 1 \label{5.7}
\end{eqnarray}
while the variance of phase $(\Delta \Phi_{PB})^2$ in the same state within the PB formalism has the form 
\begin{equation}
(\Delta \Phi_{PB})^2 = \frac{\pi^2}{3}, \quad  \mathrm{for}\,n=0,1,\dots
\label{5.8}
\end{equation}
Therefore, the probability distribution of the phase for the GW operator in any number eigenstate $\ket{\underline{n}}$ is not uniform. In contrast, the respective probability distribution calculated within the PB approach is uniform as it is expected for the number eigenstates.  Hereby, at least, from this point of view the PB formalism seems to be more adequate then the GW one. 
Also, the existing experiments on the fluctuations of phase in coherent photon states seem to confirm the PB theory \cite{34,35,36,37,38,39}. Nevertheless, the results are still under discussion and in fact, can hardly be considered as conclusive \cite{40,41}. Consequently, yet other approaches to the problem of defining phase operator are possible. 
Here we arrive at the point where the generalized Weyl quantization rule on $S^1\times\mathbb{R}^1$ described in the proceeding sections can be applied.   A quick glance at the formulas (\ref{4.7}) and (\ref{5.1}) is sufficient to note   that the GW phase operator (\ref{5.1}) with $\Phi_0=-\pi$ can be considered as the following projection of $-\hat{\Theta}$
\begin{equation}
\hat{\Phi} =\hat{\Pi}\cdot \left( -\hat{\Theta} \right) \cdot \hat{\Pi} \label{5.9}
\end{equation}
where
 \begin{equation}
\hat{\Pi} : \left\{
\begin{array}{ll}
L^2(S^1) \ni \ket{n}\mapsto \ket{\underline{n}}\in L^2(\mathbb{R}^1)&,n=0,1,2,\dots,\\
L^2(S^1) \ni \ket{n}\mapsto 0\in L^2(\mathbb{R}^1) &, n=-1,-2,\dots
\end{array} \right. \label{5.10}
\end{equation}
and we write briefly $\hat{\Phi}=\hat{\Phi}_{GW}$ for $\Phi_0 = -\pi$
Formulae (\ref{5.9}) and (\ref{5.10}) suggest the following approach to the quantum phase conundrum
\begin{enumerate}
\item[(1)]{} First, we embed the Hilbert space $L^2(\mathbb{R}^1)$ in $L^2(S^1)$ by
\begin{equation}
J:L^2(\mathbb{R}^1)\ni \sum_{n=0}^{\infty} c_n\ket{\underline{n}} \mapsto \sum_{n=0}^{\infty} c_n\ket{n} \in L^2(S^1) \label{5.11}.
\end{equation}
\item[(2)]{} Then for any classical observable relevant to the phase $f=f(\Phi)$ we assign its quantum counterpart in a state $ \ket{\underline{\Psi}}\in L^2(\mathbb{R}^1)$ by quantizing the classical observable on the circle $f=f(-\Theta)$ in the state $J(\ket{\underline{\Psi}}) \in L^2(S^1)$ according to the generalized Weyl quantization rule. 
\item[(3)]{} Finally, to find any measurable quantity relevant to phase in a state $ \ket{\underline{\Psi}}\in L^2(\mathbb{R}^1)$ we calculate the respective measurable quantity on the circle in the state $J(\ket{\underline{\Psi}}) \in L^2(S^1)$.
\end{enumerate}
To be more precise, given $f=f(\Phi)$ we substitute $f(-\Theta)$ into (\ref{3.16}). We assume that (\ref{2.10}) holds true, so the result is independent of the kernel $\mathcal{K}$. Consequently, we omit the symbol $[\mathcal{K}]$ and we obtain 
\begin{eqnarray}
\braket{j|\hat{f}|k} &=& \int_{-\pi}^{\pi} f(-\Theta) \exp{\{-i(j-k)\Theta \}}\frac{\mathrm{d}\,\Theta}{2\pi} = \nonumber \\ &=& 
\int_{-\pi}^{\pi} f(\Phi) \exp{\{i(j-k)\Phi \}}\frac{\mathrm{d}\,\Phi}{2\pi} \label{5.12}
\end{eqnarray}
Therefore
\begin{equation}
\hat{f} = \sum_{j,k=-\infty}^{\infty} \left[   \int_{-\pi}^{\pi} f(\Phi) \exp{\{i(j-k)\Phi \}}\frac{\mathrm{d}\,\Phi}{2\pi} \right] \ket{j}\bra{k}.
\label{5.13}
\end{equation}
This is our operator acting in $L^2(S^1)$ corresponding to the observable $f(\Phi)$. Afterwards, the expectation value of $f(\Phi)$ in a state $\ket{\underline{\Psi}} \in L^2(\mathbb{R}^1)$, $\braket{\underline{\Psi}|\underline{\Psi}}=1$, can be found from the rule ($3$) as follows
\begin{eqnarray}
\braket{f(\phi)}_{\ket{\underline{\Psi}}} &=& \braket{ {\underline{\Psi}}| (\hat{\Pi} \hat{f} \hat{\Pi} ) | {\underline{\Psi}}} = \nonumber \\
&=& \int_{-\pi}^{\pi} f(\phi) \bra{\underline{\Psi}}\frac{1}{2\pi} \sum_{j,k=0}^{\infty} \exp{\{ i(j-k) \phi \} } \ket{\underline{j}} \bra{\underline{k}} \ket{\underline{\Psi}} \mathrm{d}\phi \label{5.14}.
\end{eqnarray}
Let $\mathcal{B}([-\pi,\pi))$ be the family of Borel sets on $[-\pi,\pi)$ and $\mathcal{L}_+(L^2(\mathbb{R}^1)) $ denote the set of bounded positive operators on $L^2(\mathbb{R}^1)$, then the map
\begin{equation}
M_0: \mathcal{B}([-\pi,\pi)) \ni X \mapsto \frac{1}{2\pi} \sum_{j,k=0}^{\infty}\int_{X} \exp{\{ i(j-k) \phi\}} \mathrm{d}\phi \ket{\underline{j}}\bra{\underline{k}} \in \mathcal{L}_+(L^2(\mathbb{R}^1))
\label{5.15}
\end{equation}
defines a \textit{positive operator valued (POV) measure} on $[-\pi, \pi)$. 
This is precisely the POV measure defined by Shapiro and Shepard \cite{19} and by Busch, Grabowski and Lahti \cite{13} as the quantum representation of the phase. Thus we arrive at the conclusion that our approach to the quantum phase based on generalized Weyl quantization on the phase space $S^1\times \mathbb{R}^1$ is equivalent to the POV measure formalism given in \cite{13}. In consequence the projection $\hat{\Pi}$ defined by (\ref{5.10}) is the Naimark projection, the POV measure (\ref{5.15}) is a \textit{compression} of the spectral measure $E$  
\begin{equation}
E: \mathcal{B}([-\pi,\pi)) \ni X \mapsto \frac{1}{2\pi} \sum_{j,k=-\infty}^{\infty}\int_{X} \exp{\{ i(j-k) \Theta\}} \mathrm{d}\Theta \ket{ {j}}\bra{ {k}} \in \mathcal{L}_+(L^2(S^1))
\label{5.16}
\end{equation}
to $L^2(\mathbb{R}^1)$
and $E$ is a \textit{dilation} of $M_0$ \cite{42,43} ($E$ is also called the \textit{projective valued  (PV) measure}).
Now, we are going to show that our approach described by the points $(1)$, $(2)$ and $(3)$ is also equivalent to the Pegg-Barnett approach. To this end consider an observable $f=f(\Phi)$ relevant to the phase. If $\ket{\underline{\Psi}} \in \mathrm{span} \{ \ket{\mathrm{n}} \}_{n=0}^{\infty}$, $\braket{\underline{\Psi}|\underline{\Psi}}=1$, then according to the PB formalism the expectation value of the respective quantum observable  $f(\hat{\Phi}_{PB})$ in the state $\ket{\underline{\Psi}}$ is given by (see (\ref{5.4}),(\ref{5.5}) and (\ref{5.6}))
\begin{eqnarray}
&&\braket{f(\hat{\Phi}_{PB})}_{\ket{\underline{\Psi}}}= \lim_{s\to\infty} \braket{{\underline{\Psi}}|f(\hat{\Phi}_{PB}^{(s)})|{\underline{\Psi}}}=  \lim_{s\to\infty} \sum_{m=0}^{s} f(\Phi_m) \left|\braket{\underline{\Psi}| \Phi_m} \right|^2  
 =\lim_{s\to\infty} \frac{1}{s+1}  \nonumber   \\  \sum_{m=0}^{s}
&& \left[ f\left( \Phi_0 +\frac{2m\pi}{s+1} \right) \sum_{j,k=0}^{s} \exp{ \bigg\{ i(j-k) \left(\Phi_0 +\frac{2m\pi}{s+1}\right)\bigg\}}    \braket{\underline{\Psi} | \underline{j}} \braket{\underline{k}|\underline{\Psi}  }
\right]    
\end{eqnarray}
But the last expression can be rewritten as the integral and finally we have 
\begin{equation}
\braket{f(\hat{\Phi}_{PB})}_{\ket{\underline{\Psi}}} = \int_{\Phi_0}^{\Phi_0+2\pi } f(\Phi) \bra{\underline{\Psi}}
 \sum_{j,k=0}^{\infty} \exp{  \{ i(j-k)  \Phi  \}} \ket{\underline{j}} \braket{\underline{k}|\underline{\Psi}  } \mathrm{d} \Phi
\end{equation}
for any $\ket{\underline{\Psi}} \in \mathrm{span}\{ \ket{\underline{n}} \}_{n=0}^{\infty}$, $\braket{\underline{\Psi}|\underline{\Psi}}=1$. So it is also true for any $\ket{\underline{\Psi}}\in L^2(\mathbb{R}^1)$
and choosing $\Phi_0=-\pi$ one arrives at the formula $(\ref{5.14})$. This ends the proof (see also the proof in \cite{19}) and we conclude that the PB approach to the problem of defining the quantum phase is equivalent to the POV measure approach \cite{13,19} and to the approach defined by our points $(1)$, $(2)$ and $(3)$ in which quantum phase observables are given by the self-adjoint operators on $L^2(S^1)$.


\section{Concluding remarks}\label{sec6}
We have developed the formalism pertinent to the generalized Weyl quantization on the cylindrical phase space $S^1\times \mathbb{R}^1$. Next we have shown that quantum physical quantities relevant to the phase can be represented by the self-adjoint operators on $L^2(S^1)$. It has been proved that this approach to the problem of defining quantum phase is equivalent to the POV measure approach proposed in \cite{13,19} and to the famous Pegg-Barnett approach \cite{16,17,18}. Our approach reveals the fact that in the POV measure formalism for describing the quantum space the respective Naimark extension of the Hilbert space  $L^2(\mathbb{R}^1)$ is the Hilbert space $L^2(S^1)$. Now, since the number operator $\hat{N}$ can be considered as a Naimark projection of $\frac{1}{\hslash}\hat{L}$ 
\begin{equation}
\hat{N} = \hat{\Pi} \cdot \left( \frac{1}{\hslash} \hat{L} \right) \cdot \hat{\Pi} \label{6.1} 
\end{equation}
one may expect that any quantum observable describing the physical quantity depending on phase and number of photons can be represented by the appropriate self-adjoint operator in $L^2(S^1)$ obtained by the generalized Weyl quantization rule developed in section \ref{sec2}.  
In particular it is known that the uncertainty principle for $\hat{\Theta}$ nad $\hat{L}$ in the state $\Psi = \Psi(\Theta) \in L^2(S^1)$, $\int_{-\pi}^{\pi} \Psi^*(\Theta)\Psi(\Theta) \mathrm{d} \Theta =1$,
$\int_{-\pi}^{\pi} \Psi^*(\Theta)\Theta\Psi(\Theta) \mathrm{d} \Theta =0$, reads \cite{44,45}
\begin{equation}
\Delta\Theta \Delta L \geq \frac{1}{2}\hslash \left| 1-2\pi |\Psi(\pi)|^2 \right| \label{6.2}.
\end{equation} 
 Assume that $\Psi(\Theta) \in J(L^2(\mathbb{R}^1))$. Then 
\begin{equation}
J(L^2(\mathbb{R}^1))\ni \Psi(\Theta) = \sum_{n=0}^{+\infty} c_n  \frac{1}{\sqrt{2\pi}} \exp{\{ in\Theta \}} \Rightarrow \Psi(\pi) =  \sum_{n=0}^{+\infty} c_n  \frac{(-1)^n c_n}{\sqrt{2\pi}}
\label{6.3}
\end{equation}
and one should expect that the uncertainty relation for the phase $\hat{\Phi}$ and the number operator $\hat{N}$ in the state $L^2 (\mathbb{R}^1) \ni \ket{\underline{\Psi}}=\sum_{n=0}^{+\infty} c_n \ket{\underline{n}}$ has the form
\begin{equation}
\Delta \Phi \Delta N \geq \frac{1}{2} \left| 1- |\sum_{n=0}^{+\infty}(-1)^n c_n|^2 \right| \label{6.4}.
\end{equation}
This problem will be investigated elsewhere. 
 
{\bf Acknowledgments}
 
M. P. was partially supported by the CONACYT (Mexico) grant no.~103478.
 

\end{document}